# Molecular Dynamics Modeling of Epoxy Resins using the Reactive Interface Force Field

Gregory M. Odegard[1]\*, Sagar U. Patil[1], Prathamesh P. Deshpande[1]
Krishan Kanhaiya[2], Jordan J. Winetrout[2], Hendrik Heinz[2]
Sagar P. Shah[3], Marianna Maiaru[3]

[1]*Michigan Technological University, Houghton, MI 49931 USA*
[2]*University of Colorado at Boulder, Boulder, CO 80309 USA*
[3]*University of Massachusetts Lowell, Lowell, MA 01854 USA*

*Corresponding author email: gmodegar@mtu.edu*

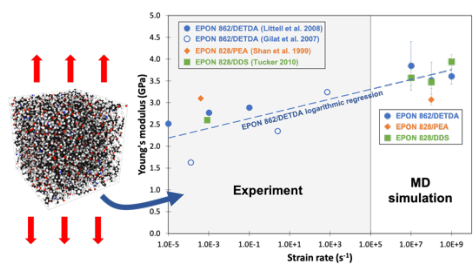

**Abstract**

Predictive computational modeling of polymer materials is necessary for the efficient design of composite materials and the corresponding processing methods. Molecular dynamics (MD) modeling is especially important for establishing accurate processing-structure-property relationships for neat resins. For MD modeling of amorphous polymer materials, an accurate force field is fundamental to reliable prediction of properties. Reactive force fields, in which chemical bonds can be formed or broken, offer further capability in predicting the mechanical behavior of amorphous polymers subjected to relatively large deformations. To this end, the Reactive Interface Force Field (IFF-R) has been recently developed to provide an efficient means to predict the behavior of materials under these conditions. Although IFF-R has been proven to be consistent for some crystalline organic and inorganic systems, it has not yet been proven to be consistent for amorphous polymer systems. The objective of this study is to use IFF-R to predict the thermo-mechanical properties of three different epoxy systems and validate with experimental measurements. The results indicate that IFF-R predicts thermo-mechanical properties that are consistent with experiment. Therefore, IFF-R can be used to reliably establish mechanical properties of polymers on the molecular level for future design of new composite materials and processing methods.

## 1. Introduction

Epoxy-based composite materials are the primary structural material used in most modern commercial and military aircraft. Their high specific strength and specific modulus make them ideal for this purpose. There is strong interest in developing higher-performing fiber/epoxy composites for improved durability for future aircraft and for reduced weight for crewed deep-space vehicles. Predictive computational modeling can be used to greatly facilitate the development of new epoxy resins and epoxy composites.

Molecular Dynamics (MD) simulation is a powerful tool for predicting the effect of molecular structure on thermo-mechanical properties of epoxy resins. A key component of MD simulation is the selection of an appropriate force field to describe the interaction between the atoms in an MD model. Force fields contain a series of energy terms associated with different degrees of freedom in a discrete molecular system, including the bonded terms (bonds, angles, dihedrals) and non-bonded terms (electrostatic interactions, van der Waals interactions, hydrogen bonding). Over the years, force fields have evolved greatly. The Class I force fields, such as CHARMM[1], AMBER[2], DREDING[3], and OPLS[4], contain simple bonded terms that are typically harmonic and uncoupled. Class II force fields, such as COMPASS[5], MM2[6], CFF[7], and MMFF[8]; are computationally more expensive, but include higher-order bonded energy terms and additional bonded cross-interaction terms (e.g. bond-angle, bond-dihedral, angle-dihedral) for improved accuracy.

Reactive force fields are particularly useful for simulating systems that undergo large mechanical deformations and/or have chemical reactions. Perhaps the most well-known reactive force field is ReaxFF[9]. ReaxFF is well-proven to predict chemical reactivities and material properties under a wide range of conditions for various chemical species. However, the powerful strengths of ReaxFF come at a cost. MD simulations using ReaxFF are computationally demanding and separate parameter sets must be established to simulate specific conditions and sets of elements. That is, there is not a single ReaxFF parameter set that can be used as a workhorse for a wide range of material systems. For polymers, this is a particular problem because engineering resins contain a wide range of atom types (C, H, O, N, S, F, Si, etc.). Currently, there is no single ReaxFF parameter set that can be generally used for all polymer systems, although the Liu et al parameter set[10] has been shown to work well for predicting the mechanical response of epoxies[11, 12], epoxy-based composites[13, 14], and PEEK[15]; and the parameter set of Damirchi et al.[16] has been shown to simulate epoxy/CNT interaction accurately.

The Interface Force Field (IFF)[17] is a unique force field that is built on various Class I and Class II force fields (such as the PCFF force field[18]) to accurately simulate a wide range of materials and material interfaces[19-23]. It allows the simulation of various materials types, such as metals, ceramics, polymers, and biomacromolecules at the same time, and in higher accuracy than available before. Broad applicability and higher accuracy are related to order-of-magnitude improved representations of chemical bonding, interpretability of all parameters, and systematic validation of structures and energies. Recently, a modified version of IFF has been developed to allow the simulation of bond dissociation[24]. This so-called Reactive Interface Force Field (IFF-R) uses Morse potentials to describe covalent bond interactions instead of traditional fixed-bond harmonic potentials. Thus, IFF-R can effectively simulate covalent bond disassociation associated with large mechanical deformations. IFF-R has been proven to work well for predicting physical and mechanical properties of carbon nanotubes and crystalline polyacrylonitrile, cellulose, and FCC iron[24]. However, it has not yet been explored for high-performance thermosetting resins with an amorphous molecular structure.

The objective of this paper is to demonstrate the consistency of IFF-R for three resins with different sets of epoxide and amine monomers. Predicted thermo-mechanical properties using IFF-R are compared to the corresponding experimental values. Using simple strain rate scaling for mechanical properties, it is shown that IFF-R predicts mass density, elastic properties, yield strength, glass transition temperature, and thermal expansivity that are consistent with experiments. In this paper, the modelled materials are described first, followed by an explanation of the MD modeling protocols. The experimental methods are described next, and a comparison of the predictions and measurements follows.

## 2. Materials

Three epoxy polymer systems were considered in this study. These three systems are described below, as indicated by

- EPON 862/EPIKURE W: Diglycidyl ether bisphenol F (DGEBF) with a diethyltoluenediamine (DETDA) curing agent
- EPON 828/Jeffamine D-230: Diglycidyl ether bisphenol A (DGEBA) with a polyetheramine (PEA) curing agent
- EPON 828/OMICURE 33-DDS: DGEBA with a 3,3' diaminodiphenyl sulfone (33DDS) curing agent

The molecular structures of the EPON 862/EPIKURE W (henceforth referred to as EPON 862/DETDA), EPON 828/Jeffamine D-230 (henceforth referred to as EPON 828/PEA), and EPON 828/OMICURE 33-DDS (henceforth referred to as EPON 828/DDS) systems are shown in Figures 1, 2, and 3, respectively. These systems were chosen because they are both highly benchmarked. The selection of these three resins also provides a contrast between aromatic and aliphatic hardeners. For the Jeffamine D230 molecule, a mixture of $n = 2$ and $n = 3$ molecules (Figure 2) were simulated such that the average value of $n$ was 2.5.

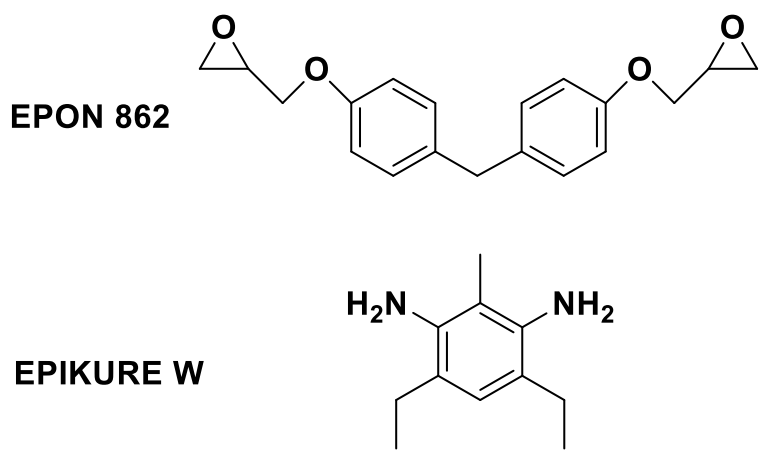

*Figure 1 – EPON 862/DETDA molecular structure*

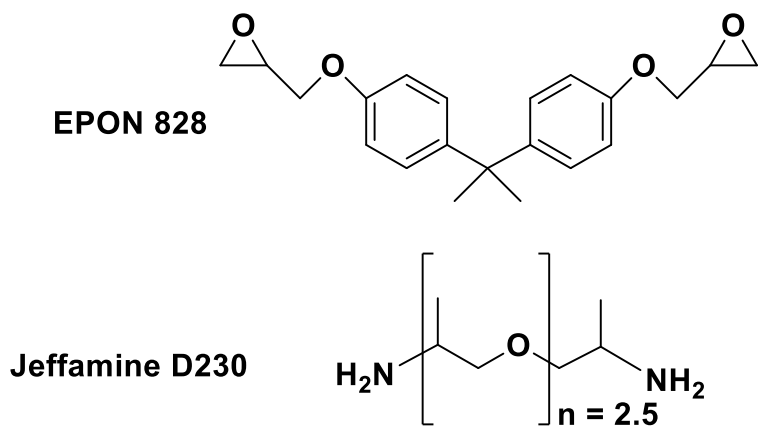

*Figure 2 – EPON 828/PEA molecular structure*

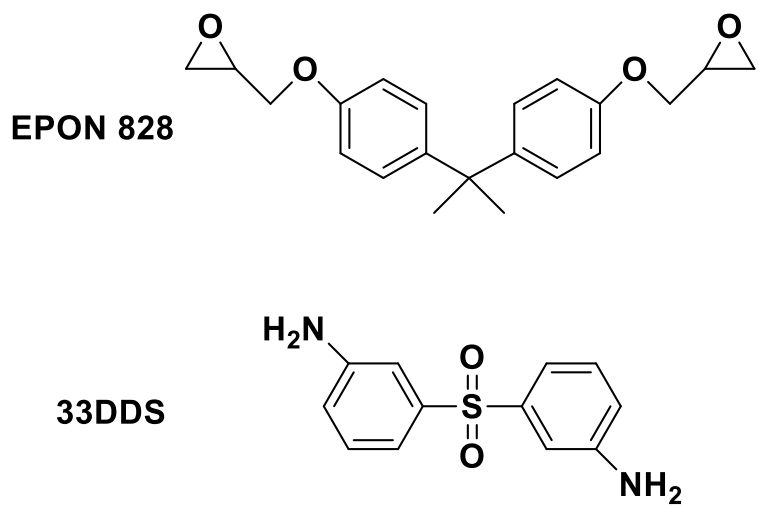

*Figure 3 – EPON 828/DDS molecular structure*

## 3. MD simulation details

The LAMMPS (Large-scale Atomic/Molecular Massively Parallel Simulator) software package was used for all MD simulations described herein[25], utilizing the IFF-R force field. The virtual π orbitals that have been used previously with IFF[26] were not used in the current study because no aromatic reinforcement surfaces were simulated (e.g. graphene or carbon nanotubes). The Lennard-Jones diameters were chosen per guidelines described by Heinz et al. [17]. Nose-Hoover algorithms were used for both the thermostats and barostats for all of the simulations discussed herein[27-29]. The MD modeling algorithm consisted of three stages: model building, crosslinking, and property prediction.

In the first stage, the MD models were assembled, densified, annealed, and equilibrated to their equilibrium densities at room temperature. The models were initially built with the monomers in a low-density mixture, which was gradually compressed to a target mass density at room temperature. After reaching the target density, the models were annealed in the NVT (controlled volume and temperature) ensemble by ramping up to an elevated temperature immediately followed by a ramp down to room temperature. The annealing process was followed by an equilibration at room temperature in the NPT (constant pressure and temperature) ensemble. Table 1 shows the simulation parameters for all three epoxy systems. Replicates of both systems were built for statistical sampling purposes, which is also listed in Table 1. Unless noted otherwise, the pressures for all NPT simulations were all set to 1 atm.

*Table 1 - MD simulation parameters for model building, densification, annealing, and equilibration*

| Simulation parameter | EPON 862/DETDA | EPON 828/PEA | EPON 828/DDS |
|---|---|---|---|
| Epoxy/crosslinker monomers | 90/45 | 304/152 | 170/85 |
| Total number of atoms | 5,265 | 21,256 | 10,795 |
| Target mass density | 1.17 g/cc | 1.16 g/cc | 1.17 g/cc |
| Replicates | 5 | 5 | 5 |
| Densification temperature | 300 K | 300 K | 300 K |
| Densification simulation time (time steps) | 8 ns (1 fs) | 4 ns (1 fs) | 4 ns (0.1 fs) |
| Annealing temperature | 500 K | 600 K | 500 K |
| Annealing ramp rate | 20 K/ns | 50 K/ns | 20 K/ns |
| Equilibration temperature | 300 K | 300 K | 300 K |
| Equilibration time (time steps) | 1 ns (1 fs) | 1.5 ns (1 fs) | 2 ns (1 fs) |

In the second stage, the models were crosslinked using the "fix bond/react" command[30] in LAMMPS to the maximum crosslink density possible, where the crosslink density is defined as the ratio of the total number of covalent bonds actually formed to the total number of covalent bonds that could potentially be formed. As discussed previously[12], it is unrealistic to achieve crosslink densities of 100% for epoxies. The simulation conditions used for the crosslinking process for each epoxy system are provided in Table 2. After crosslinking, an additional annealing step was imposed on the EPON 828/PEA system (temperature ramp-down between 600 and 300 K at 50 K/ns). For all three systems, an NPT simulation was used to obtain the equilibrated mass density after crosslinking. The parameters for this equilibration are provided in Table 2. Figure 4 shows a representative simulation box for the EPON 828/PEA system.

*Table 2 - MD simulation parameters for model crosslinking and final equilibration*

| Simulation parameter | EPON 862/DETDA | EPON 828/PEA | EPON 828/DDS |
|---|---|---|---|
| Crosslinking simulation time (time steps) | 6 ns (0.1 fs) | 6.5 ns (1 fs) | 3 ns (0.1 fs) |
| Crosslinking simulation temperature | 450 K | 400 K | 800 K |
| Average crosslink density | 92% | 92% | 93% |
| Equilibration simulation time (time steps) | 1 ns (1 fs) | 2 ns (1 fs) | 1 ns (1 fs) |
| Equilibration temperature | 300 K | 300 K | 300 K |

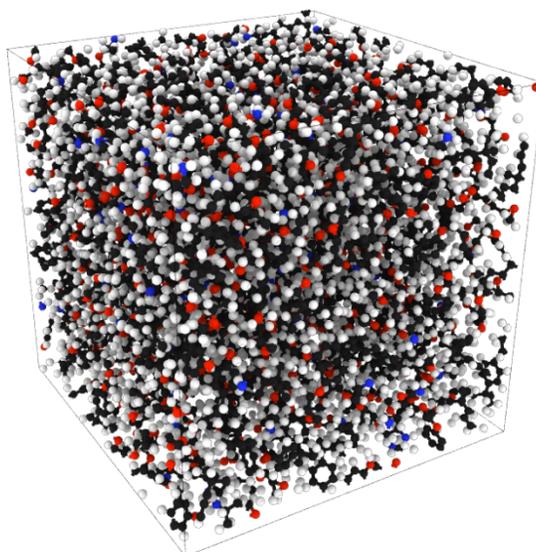

*Figure 4 – Representative MD model of EPON 828/PEA*

In the third stage of MD simulation, the crosslinked systems were subjected to mechanical and thermal loads to predict Young's modulus, yield strength, glass transition temperature ($T_g$), and the coefficient of linear thermal expansion (CTE). For the thermal properties, each MD model was subjected to steady heating to an elevated temperature and subsequent cooling to room temperature. The NPT ensemble was used to observe the density and volume over the entire temperature range. The density-temperature relationship was fitted with a bilinear regression model using the "segmented" package in R[31]. From an initial estimate, the optimal breakpoint was calculated, which was taken to be the $T_g$, as shown with a representative case in Figure 5. Additionally, the volume-temperature ($V$-$T$) plot was fit with a cubic polynomial regression model above and below $T_g$ to obtain the CTE at constant pressure, as given by

$$\alpha = \frac{1}{3V}\left(\frac{dV}{dT}\right)_p \tag{1}$$

Figure 5 also shows a representative volume-temperature graph as well as the slope of the $V$-$T$ curves above and below $T_g$. The MD parameters associated with the thermal property calculations for each polymer system are given in Table 3.

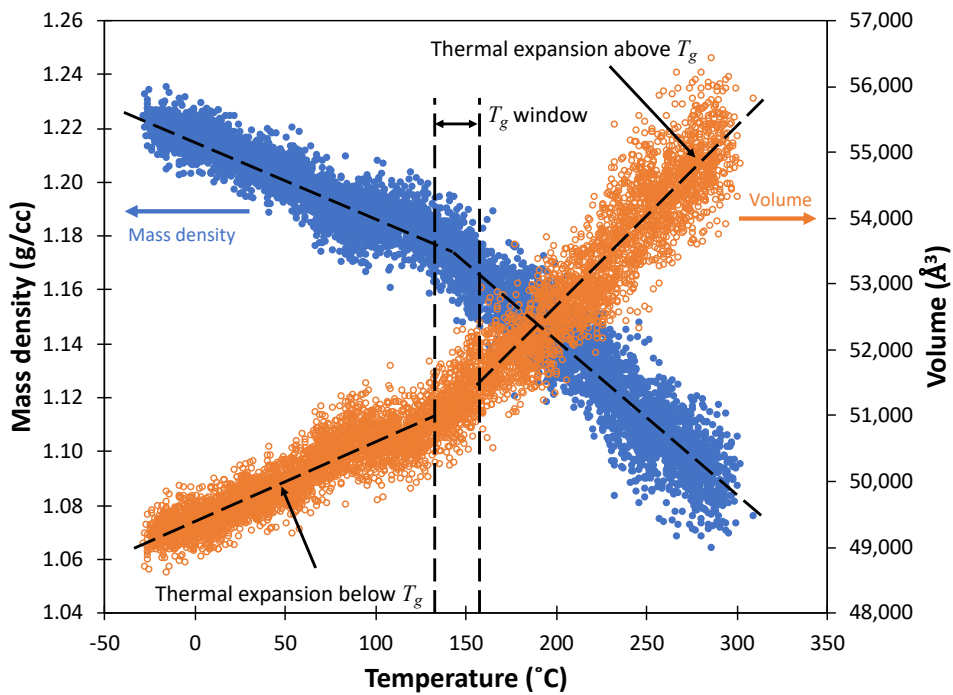

*Figure 5 – Density and Volume of a representative EPON 862/DETDA MD model as a function of temperature*

*Table 3 - MD simulation parameters for thermal and mechanical property calculations*

| Simulation parameter | EPON 862/DETDA | EPON 828/PEA | EPON 828/DDS |
|---|---|---|---|
| Temperature range | 250 – 600 K | 100 – 600 K | 250 – 600 K |
| Heating rate | 50 K/ns | 50 K/ns | 50 K/ns |
| Cooling rate | 50 K/ns | 50 K/ns | 50 K/ns |
| Shear deformation | 10% | 20% | 10% |
| Shear deformation strain rate (time steps) | $2\times10^7$ s$^{-1}$ (1 fs) $2\times10^8$ s$^{-1}$ (1 fs) $2\times10^9$ s$^{-1}$ (1 fs) | $2\times10^7$ s$^{-1}$ (1 fs) $2\times10^8$ s$^{-1}$ (1 fs) $2\times10^9$ s$^{-1}$ (1 fs) | $2\times10^7$ s$^{-1}$ (1 fs) $2\times10^8$ s$^{-1}$ (1 fs) $2\times10^9$ s$^{-1}$ (1 fs) |

To determine the bulk modulus ($K$), the MD models were subjected to an elevated pressure (5000 atm) at room temperature (NPT ensemble), and the corresponding equilibrium volume was compared to that from the ambient pressure (1 atm) at room temperature. The bulk modulus was subsequently calculated as described in detail elsewhere[32]. To determine the shear moduli ($G$), shearing deformations were performed in the *yz, xy, xz* planes[14] at 300 K and three different strain rates (for the EPON 862/DETDA and EPON 828/DDS systems) using the simulation parameters provided in Table 4. Figure 6 shows a representative shear stress-shear strain curve for the EPON 862/DETDA system at a strain rate of $2\times10^8$ s[-1]. For the stress-strain curve associated with each replicate, shearing plane, and strain rate, the bilinear breakpoint was determined by observing the strain at which the slope changed significantly. The shear modulus was calculated as the slope of the line before the breakpoint. The Young's moduli ($E$) and Poisson's ratios ($\nu$) for each model were determined from the corresponding values of bulk modulus and the average shear modulus using standard isotopic elasticity equations[33]

$$E = \frac{9KG}{3K+G} \quad (2)$$

$$\nu = \frac{3K-2G}{2\left(3K+G\right)} \quad (3)$$

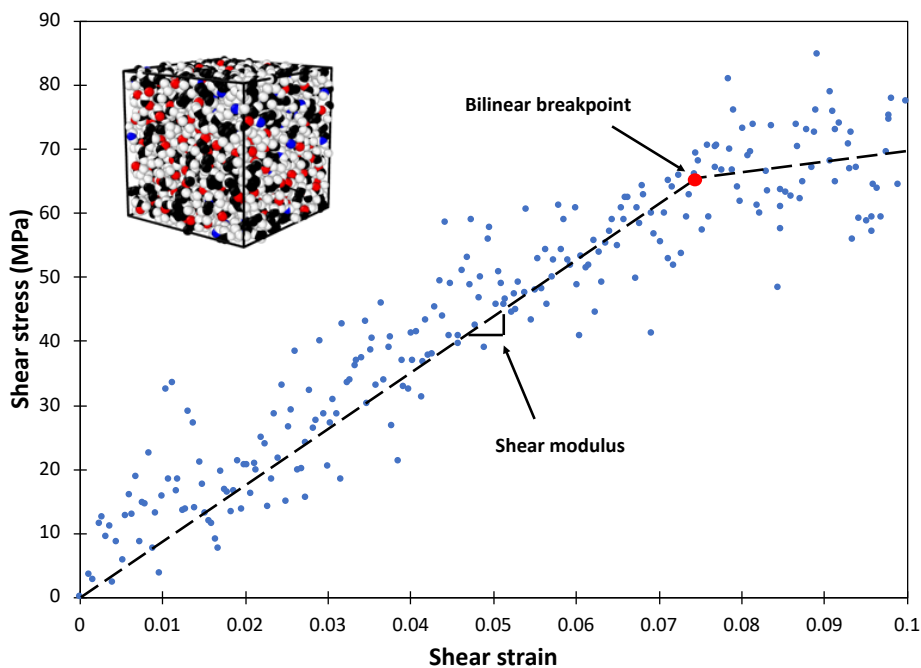

*Figure 6 – Representative shear stress/strain curve for EPON 862/DETDA*

For the uniaxial yield strength, the von Mises stress was determined from the individual stress components during the shear deformations:

$$\sigma_{vM} = \sqrt{\frac{1}{2}\left[\left(\sigma_x - \sigma_y\right)^2 + \left(\sigma_y - \sigma_z\right)^2 + \left(\sigma_z - \sigma_x\right)^2 + 6\left(\tau_{xy}^2 + \tau_{xz}^2 + \tau_{yz}^2\right)\right]} \quad (4)$$

The corresponding yield strength was the von Mises stress at the same breakpoints determined for the shear modulus, described above. Thus, the yield strength was determined for each replicate, shearing plane, and strain rate. Figure 7 shows a representative von Mises/shear strain curve for the EPON 862/DETDA system at a strain rate of $2\times10^8$ s$^{-1}$, with the breakpoint and yield strength value shown. The overall procedure for comparing the computationally-derived bilinear breakpoint with the laboratory length-scale yield stress is consistent with previous work[11]. It has been previously shown that for polymers at this length scale, observed yield is typically characterized by chain segment configurational changes[34-36], which is classically described by the Argon theory[37, 38].

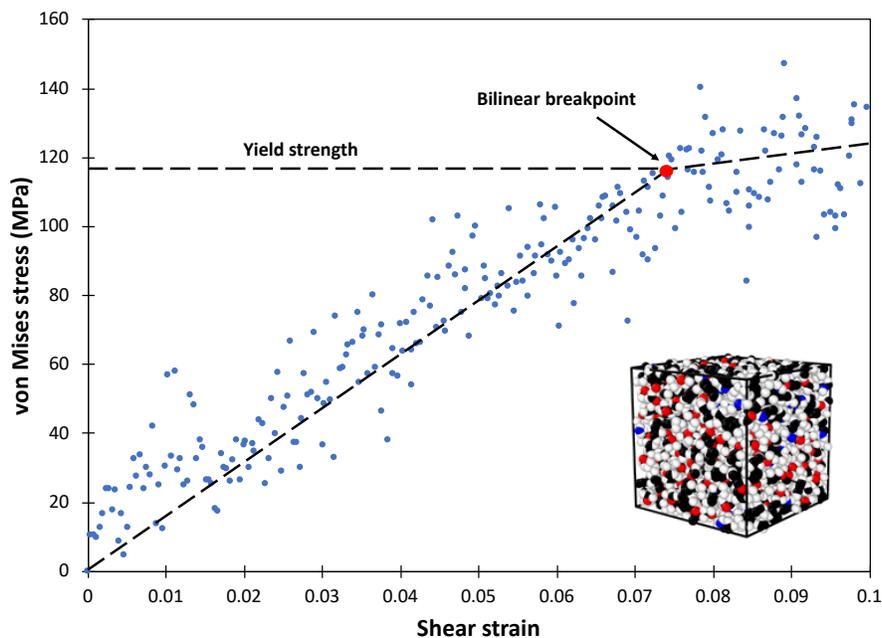

*Figure 7 – Representative von Mises/shear strain curve for EPON 862/DETDA*

## 4. Cooling rate effect on $T_g$

It is important to note that no cooling rate correction factors were used for the $T_g$ predictions performed herein, as have been employed previously[39-41] for MD predictions of $T_g$. It is well-known that cooling rates affect the experimental measurement of $T_g$[42-46]. Under laboratory conditions, rapid cooling through the $T_g$ window does not provide the polymer molecular structure enough time to respond (via chain segment configurational changes) to rapid drops in free volume. This essentially "locks in" a non-equilibrated molecular configuration that contains different levels of free volume relative to a fully equilibrated system[44-46]. This same effect is generally not observed for heating-related $T_g$ measurements[44]. This section discusses this phenomenon and the decision to not use correction factors for the $T_g$ predictions in this study.

Amorphous polymer systems contain finite amounts of free volume, that is, volume pockets that do not contain any mass (a more rigorous definition is provided elsewhere[44]). At equilibrium, the amount of free volume in an amorphous polymer is dependent on the temperature of the system.

In general, higher temperatures correspond to higher levels of free volume, due to increasing levels of atomic motion and repulsion, and thus thermal expansion. For a given state of temperature and free volume, there corresponds an equilibrium state of polymer chain segment configurations. It is important to note that reaching such equilibrium states at temperature below $T_g$ can take significant amounts of time (days, weeks, years), whereas above $T_g$, reaching equilibrium states occurs much quicker (minutes, hours)[44].

As temperatures change, thermodynamic drivers for chain configurational changes activate. However, even though changes in temperature can be carefully controlled, the corresponding equilibrium chain configurations can take a considerably longer period of time to occur. If the cooling rate is faster than the ability of the network to respond accordingly, then the thermal contraction of the system will lock-in chain configurations that are out of equilibrium, and changes in the free volume will not be able to keep up with the thermal contraction. The faster the cooling rate, the greater the discrepancy between thermal contractions and reductions in free volume. The greater this discrepancy, the faster the system approaches the glass transition window. Thus, higher cooling rates result in higher values of "apparent" $T_g$.

During experimental measurements of $T_g$ via cooling through the transition window, the specimen is initially held at the elevated temperature for a finite amount of time (seconds, minutes, hours) before ramping down. During this hold the polymer network approaches an equilibrium in terms of polymer configuration and free volume corresponding to the elevated hold temperature. When the cooling commences, the cooling rate effect described above activates.

During MD simulation of the cooling from the *rubbery state* (above $T_g$) to the *glassy state* (below $T_g$), this same effect of cooling rate on the apparent $T_g$ is not necessarily observed. Because of the very short time scales associated with MD (nanoseconds), the cooling rate is necessarily orders of magnitude greater than that of experiment. For the simulations discussed herein, the model creation and densification occurred at room temperature (below the $T_g$ value for both of the modeled epoxy systems). Thus, the equilibrium molecular chain segment configurations associated with the models corresponded to those at equilibrium at room temperature. Although the MD modeling process involved an annealing step and a temperature ramp-up for heating $T_g$ predictions, these excursions to elevated temperatures occurred on very short timeframes (nanoseconds), which is orders of magnitude smaller than timeframes typically associated with high-temperature structural relaxation (seconds, minutes, hours). Therefore, at the beginning of the MD cool-down simulations for $T_g$ prediction, the systems already had their glassy equilibrium state chain segment configurations (even though they had free volume levels of the rubbery state), and were thus immune to experimentally-observed cooling rate effects. The predicted $T_g$ values therefore directly relate to the $T_g$ values experimentally measured using either heating methods or relatively slow cooling rates. Thus, no cooling rate correction factors were neither needed nor used.

## 4. Strain rate effect

The measured elastic and strength properties of polymer materials are generally dependent on the applied strain rate due to their viscoelastic nature. Molecular dynamics (MD) simulations are computationally demanding and can only simulate phenomena on very small times scales (nanoseconds). If an MD simulation includes a simulated deformation of a material to predict a mechanical property, the deformation must occur on the order of nanoseconds. Under typical laboratory conditions, experimental mechanical testing occurs over much higher time scales, such as seconds, minutes, or hours. Thus, the time scales of MD simulations and experimental testing differ by several orders of magnitude, as does their corresponding strain rates. Therefore, it is expected that predicted and measured mechanical properties should also differ significantly. This discrepancy is typically referred to as the *strain rate effect*.

Odegard et al.[11] first quantified the strain rate effect for the EPON 862/DETDA system using MD techniques and the ReaxFF force field[9, 10]. MD simulations were used to predict the elastic modulus and yield strength of this epoxy system at two strain rates ($1 \times 10^8 \, \text{s}^{-1}$, $2 \times 10^8 \, \text{s}^{-1}$). These properties were compared to experimentally-measured modulus and yield strength values on the same material system reported by Littell et al.[47] at three strain rates: $1 \times 10^{-5} \, \text{s}^{-1}$, $1 \times 10^{-3} \, \text{s}^{-1}$, and $1 \times 10^{-1} \, \text{s}^{-1}$. Thus, the strain rates differed by as much as thirteen orders of magnitude. These comparisons of modulus and strength showed a significant discrepancy between measurements and predictions, but they also showed a clear scaling trend of the properties due to the strain rate effect. This observation was verified by Radue et al.[12] for the same epoxy system. In the current study, an approach similar to Odegard et al.[11] and Radue et al.[12] is taken for reporting Young's modulus and yield strength. That is, the predicted mechanical property values will be compared to the experimentally-measured values with the expectation that there will be a discrepancy due to the strain rate effect.

## 5. Experimental details

For mass density measurement specimens of the EPON 862/DETDA system, the neat resin was degassed at room temperature in a speed mixer at 900 rpm under vacuum. The curing agent was added, and the mixture was thoroughly mixed for 5 minutes and then further degassed in a vacuum chamber at room temperature under vacuum for 30 minutes. The mixture was then cured in small aluminum dishes isothermally at 177°C until fully cured and then cut into small pieces of varying masses for mass density measurements. A total of 10 specimens were fabricated. The mass density measurements used the Archimedes Principal method (ASTM D792).

Dynamic Mechanical Analysis (DMA) was used for the $T_g$ measurements of the EPON 862/DETDA system. The DMA specimens were prepared in a closed silicone mold. Both parts of the epoxy system were thoroughly mixed in a 100:26.4 parts-by-weight stoichiometric ratio. The uncured mixture was placed into an oven set to 80°C for 20 minutes. Pre-heating the resin helped reduce the viscosity for easy degassing, which was performed in a vacuum chamber for 20 minutes at room temperature. The degassed mixture was injected into the pre-heated silicon mold and allowed to fully cure at 177°C for 150 minutes. Three fully-cured DMA specimens were then demolded and sanded to average dimensions of 35×12×3 mm, conforming to the ASTM D7028-07 standard.

A TA Instruments Q800 DMA was used to determine the $T_g$ of the three specimens. During the test, a sinusoidal displacement with an oscillation frequency of 1 Hz was applied to the specimens. The specimens were subjected to temperature sweeps (ramping from 20° C to 250° C at 5° C/min) in a single cantilever beam configuration. The tan δ response of the specimens was monitored over the entire range, and the peak tan δ value was taken as the corresponding $T_g$ value.

## 6. Results

The results of MD predictions and experimental measurements are presented in this section. All uncertainty values represent the standard error of replicate simulations/measurements. Tables 4, V, and VI list the predicted mass density values for the EPON 862/DETDA, EPON 828/PEA, and EPON 828/DDS systems, respectively, along with the corresponding densities from the experiments described herein for the EPON 862/DETDA system and from the literature for the EPON 828/PEA[48, 49] and EPON 828/DDS systems[50]. The results indicate that the MD predictions with IFF-R agree closely with experiment for all three systems.

Tables 4, 5, and 6 list the predicted and experimental thermal properties of the three epoxy systems. The experimental $T_g$ results for the EPON 862 system are those described herein, while the remaining results are from the literature[50-61]. The predicted $T_g$ values are listed for both the heating and cooling cycles. The results indicate that the predicted properties generally match the experimental values closely for $T_g$, CTE below $T_g$, and CTE above $T_g$. It's important to note that the predicted heating and cooling values for $T_g$ in Tables 4, 5, and 6 are in agreement with each other. This observation supports the discussion above regarding the effect of cooling rates on the predicted thermal response of an amorphous polymer modeled with the MD procedure used herein. Because the MD systems were initially formed and densified below $T_g$, the sub-$T_g$ configurational structure is locked in and thus the general approach to $T_g$ prediction (heating vs cooling) does not affect the results.

*Table 4 – Properties for EPON 862/DETDA*

| Property | MD prediction | Experiment |
|---|---|---|
| Mass density (g/cc) | 1.204 ± 0.003 | 1.193 ± 0.001 |
| $T_g$ (°C) | 154.8 ± 9.1 (heating) 155.1 ± 3.0 (cooling) | 153.8 ± 0.3 |
| CTE below $T_g$ (×10⁻⁵ °C⁻¹) | 7.95 ± 0.33 | 6.41 [51] |
| CTE above $T_g$ (×10⁻⁵ °C⁻¹) | 16.01 ± 0.60 | 18.59 [51] |

*Table 5 – Properties for EPON 828/PEA*

| Property | MD prediction | Experiment |
|---|---|---|
| Mass density (g/cc) | 1.153 ± 0.002 | 1.156 − 1.157 [48, 49] |
| $T_g$ (°C) | 92.8 ± 3.7 (heating) 91.1 ± 2.4 (cooling) | 80 − 90 [52-54, 58] |
| CTE below $T_g$ (×10⁻⁵ °C⁻¹) | 6.32 ± 0.28 | 6.0 − 9.0 [55, 56, 59, 60] |
| CTE above $T_g$ (×10⁻⁵ °C⁻¹) | 12.20 ± 1.25 | 13.0 - 25.9 [56, 59-61] |

*Table 6 – Properties for EPON 828/DDS*

| Property | MD prediction | Experiment |
|---|---|---|
| Mass density (g/cc) | $1.226 \pm 0.002$ | 1.240 [50] |
| $T_g$ (°C) | $184.7 \pm 3.7$ (heating) $183.7 \pm 10.6$ (cooling) | $159 - 186$ [50, 57] |
| CTE below $T_g$ ($\times 10^{-5}$ °C$^{-1}$) | $6.58 \pm 0.20$ | 6.90 [57] |
| CTE above $T_g$ ($\times 10^{-5}$ °C$^{-1}$) | $16.40 \pm 0.87$ | 17.6 [57] |

Figures 8 and 9 show the predicted and experimental[47, 50, 53, 62] Young's modulus and yield strength values, respectively, of the three epoxy systems with respect to the strain rate. For the experimental values from Galat et al.[62], the published strain-rate dependent shear tests results were used for the data points in Figures 8 and 9. The shear modulus and shear stress at yield were obtained from the published stress-strain curves and analyzed with the same procedure as described elsewhere[11]. These values were subsequently converted to Young's modulus (assuming a Poisson's ratio of 0.4 [47]) and yield strength as explained above for the predicted values from MD simulation. A best-fit logarithmic regression of the EPON 862/DETDA experimental values (from both experimental references) is included on each graph and extrapolated to the simulated strain rates.

The data in Figures 8 and 9 demonstrates that the predicted properties show general agreement with the experiment considering the influence of the strain rate. It is important to note that the magnitude of discrepancy between the experiment and predictions is significantly smaller than that reported by Odegard et al[11] for the ReaxFF force field[9] with the Liu parameter set[10]. Specifically, the comparisons of modulus of Odegard et al[39] showed a predicted modulus that was about 68% higher than the measured modulus at the $1 \times 10^{-1}$ s$^{-1}$ strain rate. The predicted yield strength was about 100% greater than the measured yield strength at the same strain rate. In the current study, at the same experimental and computational strain rates, the predicted modulus is about 22% higher than the experimental modulus, and the predicted yield strength is about 33% higher. This difference in the discrepancies in predicted and measured properties from two different force fields indicates that the force constants and functional forms associated with different force fields can have a significant impact on the magnitude of the apparent strain rate effect. Because force constants are phenomenological in nature, there is no clear physical reasoning for this. Regardless, it is clear that IFF-R predicts properties that are closer to those measured experimentally than does ReaxFF with the Liu parameter set, and the apparent strain rate effect is substantially smaller.

It is also important to discuss the trendlines for the 862/DETDA experimental data shown in Figures 8 and 9. The logarithmic trendlines were used only for the purpose of showing an approximate linear trend on the logarithmic strain rate scale. That is, they have no physical significance. For the modulus trend shown in Figure 8, the logarithmic trendline demonstrates an excellent fit to both the experimental and predicted data. For the yield strength data shown in Figure 9, the trendline does not match the predicted data with the degree of accuracy found in Figure 8. This simply means that a logarithmic trendline does not completely capture the physics of the strain rate effect on the yield strength. It is close, but not perfectly accurate. Perhaps, in the future, more data can be captured and more sophisticated fitting techniques can be explored to capture the physical trends observed with yield strength over numerous orders of magnitude of strain rate.

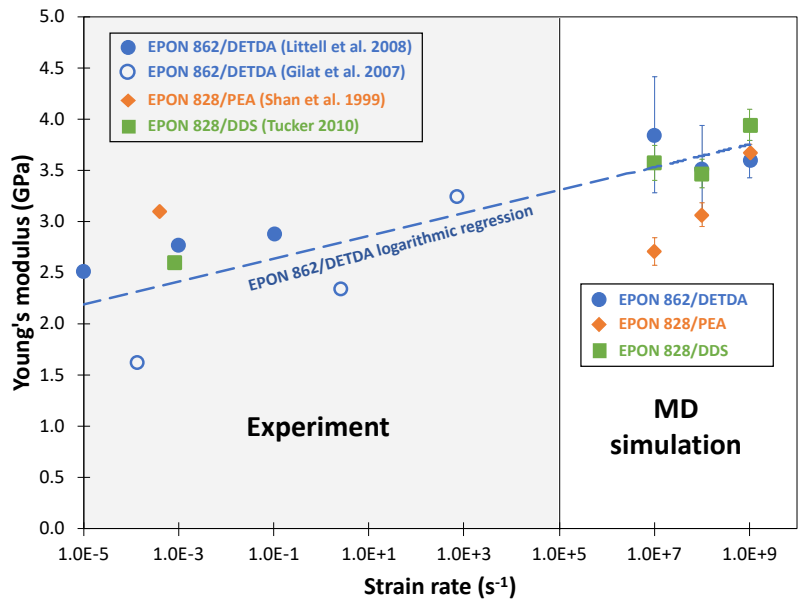

*Figure 8 – Young's modulus vs strain rate for all three epoxy systems. The EPON 862/DETDA logarithmic regression line is fit to both sets of EPON 862/DETDA experimental data*

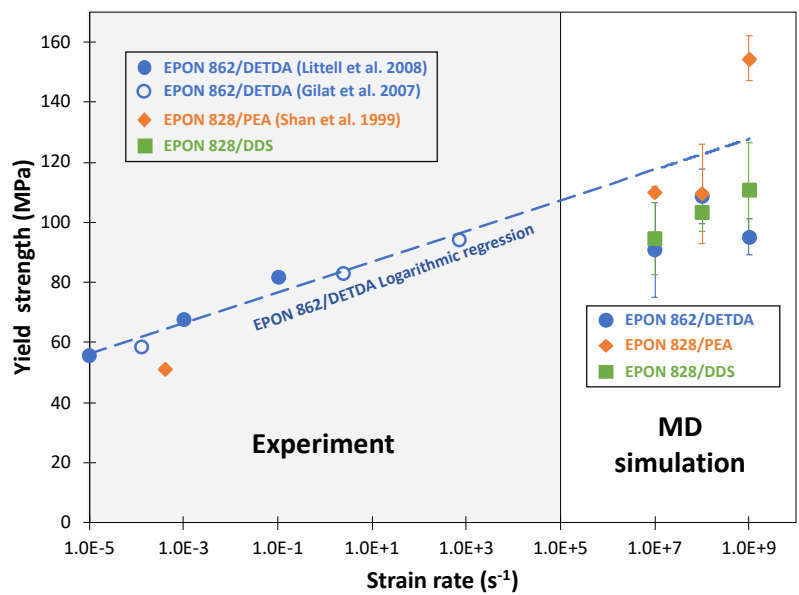

*Figure 9 – Yield strength vs strain rate for all three epoxy systems. The EPON 862/DETDA logarithmic regression line is fit to both sets of EPON 862/DETDA experimental data*

Table 7 shows a comparison of the Young's modulus predictions of the 862/DETDA system with the predictions from other MD-based studies from the literature[11, 12, 63-65]. Experimental values from the literature are also included in the table[47, 62]. Although there are numerous MD-based studies in the literature for the 862/DETDA system, only those with reasonable mass density predictions were chosen for this comparison. From the data in the table, the Class I force fields generally show the closest agreement with experiment, and do not appear to show any significant strain rate effect. The predictions with ReaxFF generally show the largest values of Young's modulus and the largest magnitude of strain-rate effect, while the IFF-R predictions are slightly smaller than the ReaxFF predictions. It is important to note that experimental measurements of Young's modulus at MD-level strain rates are not available, so the true magnitude of the strain rate effect is not known. However, the trend in the data in Figure 8 shows reasonable evidence that a significant strain rate exists, which indicates that the IFF-R and ReaxFF predictions are more reasonable than the Class I predictions.

*Table 7 – MD predicted Young's modulus for EPON 862/DETDA*

| Simulation study | Force field (type) | Predicted Young's modulus (GPa) |
|---|---|---|
| Current study | IFF-R[24] (Class II/reactive) | 3.5 – 3.8 |
| Odegard et al.[11] | ReaxFF[9, 10] (Reactive) | 4.9 – 5.1 |
| Radue et al.[12] | ReaxFF[9, 10] (Reactive) | 3.5 |
| Vashisth et al.[65] | ReaxFF[9, 66] (Reactive) | 5.0 – 6.0 |
| Kallivokas et al.[64] | DREIDING[3] (Class I) | 2.6 |
| Bandyopadhyay et al.[63] | OPLS[4] (Class I) | 2.3 |
| Experiment[47, 62] | - | 1.6 – 3.3 |

The predicted Poisson's ratios of the EPON 862/DETDA, EPON 828/PEA, and EPON 828/DDS systems were $0.40 \pm 0.01$, $0.42 \pm 0.01$, and $0.41 \pm 0.01$, respectively. The experimental value of the Poisson's ratio for the EPON 862/DETDA system is $0.40$ [47], which matches the prediction. Poisson's ratio values that approach the isotropic elastic limit (0.5) typically are not significantly affected by viscoelastic effects[67], thus they are not expected to demonstrate a significant strain rate effect.

## 7. Conclusions

In this study, the accuracy of IFF-R for amorphous polymer systems was assessed. MD simulations were used to predict the thermo-mechanical properties of three different fully cured/crosslinked epoxy systems, two with an aromatic crosslinking agent and one with an aliphatic crosslinker. The results indicate that IFF-R yields values of mass density, Young's modulus, Poisson's ratio, yield strength, $T_g$, and CTE that are consistent with experimental measurements.

The importance of these results is in the utility of IFF-R relative to other force fields used for simulating amorphous polymer systems. IFF-R incorporates the advantages of fixed-bond force fields (simulation efficiency, mathematical simplicity, physical relevance of force constants) with the advantages of reactive force fields (accurately simulating the response of covalent bonds stretched to large deformations). As a result, IFF-R is an efficient force field with force constants that can be easily accessible, and can simultaneously predict the behavior of material systems subjected to relatively large deformations. The results of this study verify that IFF-R yields predicted thermal and mechanical properties that are consistent with experiment. With this verification, IFF-R can now be used to confidently and efficiently simulate similar material systems subjected to large deformations in which the bond stretching moves beyond the harmonic regime. This is a significant advancement in the field of molecular-level polymer design and simulation.

## Acknowledgements

This research was partially supported by the NASA Space Technology Research Institute (STRI) for Ultra-Strong Composites by Computational Design (US-COMP), grant NNX17AJ32G; and NASA grant 80NSSC19K1246. SUPERIOR, a high-performance computing cluster at Michigan Technological University, was used in obtaining the MD simulation results presented in this publication.